\documentclass[prd,preprint,superscriptaddress,nofootinbib,longbibliography]{revtex4-1}
\pdfoutput=1

\usepackage{graphicx}
\usepackage{amsmath}
\usepackage{amsfonts}
\usepackage{mathtools}
\usepackage{color}
\usepackage{amssymb}
\usepackage{dcolumn}
\usepackage{bm}
\usepackage{braket}
\usepackage{microtype} 
\usepackage[linktoc=all]{hyperref}
\usepackage{cleveref}

\numberwithin{equation}{section}

\newcommand{\ud}{\mathrm{d}}

\newcommand{\bel}[1] {\begin{equation}\label{#1}}
\newcommand{\beal}[1] {\begin{eqnarray}\label{#1}}

\newcommand{\be}{\begin{equation}}
\newcommand{\bea}{\begin{eqnarray}}
\newcommand{\ee}{\end{equation}}
\newcommand{\eea}{\end{eqnarray}}

\makeatletter
\DeclareRobustCommand{\rcite}[1]{%
\rcite@aux#1,\@nil{#1}%
}
\def\rcite@aux#1,#2\@nil#3{%
\if\relax#2\relax
Ref.~\cite{#3}%
\else
Refs.~\cite{#3}%
\fi
}
\makeatother

\begin{document}
\title{The classical double copy in maximally symmetric spacetimes}

\author{Mariana Carrillo Gonz\'alez} 
\email{cmariana@sas.upenn.edu}
\affiliation{Center for Particle Cosmology, Department of Physics and Astronomy,
	University of Pennsylvania, Philadelphia, Pennsylvania 19104, USA}
\author{Riccardo Penco}
\email{rpenco@sas.upenn.edu}
\affiliation{Center for Particle Cosmology, Department of Physics and Astronomy,
	University of Pennsylvania, Philadelphia, Pennsylvania 19104, USA}
\author{Mark Trodden}
\email{trodden@upenn.edu}
\affiliation{Center for Particle Cosmology, Department of Physics and Astronomy,
	University of Pennsylvania, Philadelphia, Pennsylvania 19104, USA}

\date{\today}

\begin{abstract}
\noindent The classical double copy procedure relates classical asymptotically-flat gravitational field solutions to Yang-Mills and scalar field solutions living in Minkowski space. In this paper we extend this correspondence to maximally symmetric curved spacetimes. We consider asymptotically (A)dS spacetimes in Kerr-Schild form and construct the corresponding single and zeroth copies. In order to clarify the interpretation of these copies, we study several examples including (A)dS-Schwarzschild, (A)dS-Kerr, black strings, black branes, and waves, paying particular attention to the source terms. We find that the single and zeroth copies of stationary solutions satisfy different equations than those of wave solutions. We also consider how to obtain Einstein-Maxwell solutions using this procedure. Finally, we derive the classical single and zeroth copy of the BTZ black hole. 
\end{abstract}

\maketitle

\tableofcontents

\newpage

\section{Introduction}

In the Lagrangian formalism, gravity and Yang-Mills theories seem to be completely unrelated theories. However, if one looks at the structure of their scattering amplitudes, an unexpected relationship between them arises.  This relation is called the BCJ (Bern, Carrasco, Johannson) double copy~\cite{Bern:2008qj,Bern:2010yg,Bern:2010ue}, and at tree-level it is equivalent to the KLT (Kawai, Lewellen, Tye) relations \cite{Kawai:1985xq} between open and closed string amplitudes in the limit of large string tension. More explicitly, the BCJ double copy is a procedure that relates scattering amplitudes of different theories obtained by interchanging color and kinematic factors. This replacement is motivated by color-kinematics duality---the observation that the kinematic factors satisfy the same algebra as the color factors.

The BCJ relation holds both with and without supersymmetry, and even in the presence of spontaneous symmetry breaking \cite{Chiodaroli:2015rdg}.  The original and best-known form of the double copy is the so-called Gravity=(Yang-Mills)$^2$ relationship, relating Einstein Gravity (the ``double copy'') to two copies of Yang-Mills theory (each of which is usually referred to as a``single copy''). By replacing the kinematic factors in the amplitudes of a Yang-Mills theory with color factors, this procedure can be further extended to yield the amplitudes for a bi-adjoint scalar theory (the ``zeroth copy''). The bi-adjoint scalar is a real massless scalar with a cubic interaction $f^{abc}\tilde{f}^{ijk}\phi^{a\,i}\phi^{b\,j}\phi^{c\,k}$, where $f^{abc}$ and $\tilde{f}^{ijk}$ are the structure constants of the groups $G$ and $\tilde{G}$ respectively, and $\phi^{a\,i}$ is in the bi-adjoint representation of $G\times\tilde{G}$.

This procedure can be applied more generally to theories other than Einstein gravity and Yang-Mills, such as Born-Infield theory and the Special Galileon \cite{Cheung:2017ems,Cheung:2017pzi,Cheung:2016prv,Carrasco:2016ldy,Cachazo:2014xea,Chen:2013fya}, or  Einstein-Maxwell and Einstein-Yang-Mills theories~\cite{Nandan:2016pya,Stieberger:2015vya,Stieberger:2016lng,Chiodaroli:2014xia,Chiodaroli:2015wal,Chiodaroli:2017ngp,Chiodaroli:2015rdg}.The extent of this remarkable relation between scalar, gauge and gravity amplitudes is the subject of ongoing research, and we refer the reader to~\cite{Carrasco:2015iwa} for a more extensive review of the literature. 

The BCJ double copy is a perturbative result that has been proven at tree level \cite{Carrasco:2015iwa,Stieberger:2009hq,BjerrumBohr:2010zs,Mafra:2011kj,Cachazo:2012uq,Feng:2010my,Tye:2010dd}, but is also believed to hold at loop level~\cite{Bjerrum-Bohr:2013iza,Boels:2013bi,Carrasco:2011mn,Carrasco:2012ca,Bern:1997nh,Bern:1998ug,Bern:2013uka,Bern:2014sna,Bern:1994zx}. Recently, BCJ-like double copy relations between classical solutions have been suggested using different perturbative approaches~\cite{Saotome:2012vy,Neill:2013wsa,Luna:2016hge,Goldberger:2016iau,Goldberger:2017frp}. More broadly, this suggests the possibility of using the double copy technique to generate classical gravitational solutions from simpler, classical gauge configurations. Although this is an intrinsically perturbative procedure, it is interesting to understand the extent to which an exact relationship might persist at the classical non-perturbative level. This idea was first explored in \cite{Monteiro:2014cda}, where it was shown that a classical double copy that resembles the BCJ double copy exists for spacetimes that can be written in (single or multiple) Kerr-Schild (KS) form (see also~\cite{Luna:2015paa,Luna:2016due,Ridgway:2015fdl}). Other approaches to the classical double copy were considered in \cite{Anastasiou:2014qba,Anastasiou:2017nsz,Borsten:2015pla,Cardoso:2016amd,Cardoso:2016ngt,Chu:2016ngc}. 

Most of the work thus far has been devoted to the study of solutions in asymptotically-flat space-times. Recently, however, it has been shown that the BCJ double copy of three-point scattering amplitudes is successful also in certain curved backgrounds~\cite{Adamo:2017nia}. More precisely, it was shown that graviton amplitudes on a gravitational sandwich plane wave are the double copy of gluon amplitudes on a gauge field sandwich plane wave. It is therefore natural to wonder to what degree the classical double copy procedure can also be extended to curved backgrounds. In this paper, we take a first step in this direction by focusing on curved, maximally symmetric spacetimes. The viewpoint taken here is slightly different from the one adopted in~\cite{Adamo:2017nia}, in that there the curved background was also ``copied''.  At the level of the classical double copy, this approach would associate to the (A)dS background a single copy sourced by a constant charge filling all space. Here, on the other hand, we will treat the curved (A)dS background as fixed and find the single and zeroth copy solutions in de Sitter (dS) or Anti-de Sitter (AdS) spacetimes. 

The rest of the paper is organized as follows. In Secs. \ref{bcj} and \ref{sec:classical double copy}, we briefly review the BCJ and classical double copies respectively, and their relation to each other. In Secs. \ref{Sch} and \ref{kerr}, we construct single and zeroth copies of the (A)dS-Schwarzschild and (A)dS-Kerr black hole solutions in $d\geqslant 4$ spacetime dimensions. We then show how to use the single copy obtained from these solutions to construct the full Einstein-Maxwell solution for the charged (A)dS-Reissner-Nordstrom and (A)dS-Kerr-Newman black holes (Sec. \ref{EM}).  In Sec. \ref{others}, we consider the double copy construction for black strings and black branes in (A)dS, and in Sec. \ref{waves} we turn our attention to time-dependent solutions by studying the case of waves in (A)dS. Finally, in Sec. \ref{btz} we consider the Yang-Mills and scalar copies for the BTZ (Ba\~nados, Teitelboim, Zanelli) black hole. This last example extends the classical double copy beyond the regime of applicability of the BCJ procedure for amplitudes, since in $d=3$ there are no graviton degrees of freedom to which copies could correspond. We conclude by summarizing our results and discussing future directions in Sec. \ref{dis}.
\\

\emph{Note added:} During the completion of this paper, \cite{Bahjat-Abbas:2017htu} appeared, which also considers the classical double copy in curved spacetimes. In particular, two different kinds of double copies were considered. The so-called ``Type A'' double copy consists of taking Minkowski as a base metric and mapping both the background and perturbations. Thus, it is close in spirit to the approach followed in~\cite{Adamo:2017nia}. The ``Type B'' double copy considered in \cite{Bahjat-Abbas:2017htu} instead keeps the curved background fixed, which is the same prescription that we have adopted here. However, the examples analyzed in this paper are different from the ones in \cite{Bahjat-Abbas:2017htu}. Moreover, here we have paid particular attention to obtaining the correct localized sources for the Yang-Mills and scalar copies, and in analyzing the equations of motion that these copies satisfy in $d\geq4$ for both stationary and time-dependent cases. Contrary to the solutions analyzed in \cite{Bahjat-Abbas:2017htu}, we find that all examples we have considered lead to reasonable ``Type B'' single and zeroth copies (although time-dependent solutions warrant additional study).

\section{The BCJ double copy}\label{bcj}

To set the stage, we start with a concise review of the Gravity=(Yang-Mills)$^2$ correspondence.
The central point is that it is possible to construct a gravitational scattering amplitude from the analogous object for gluons. The gluon scattering amplitudes in the BCJ form can be expressed schematically as
\begin{equation} \label{A_YM}
A_\text{YM}=\sum_i\frac{N_i\,C_i}{D_i} \ ,
\end{equation}
where the $C_i$'s are color factors, the $N_i$'s are kinematic factors in the BCJ form, and the $D_i$'s are scalar propagators. It is convenient to expand the factors in the numerator in the half-ladder basis~\cite{DelDuca:1999rs}, so that they read
\begin{equation}
C_i=\sum_{\alpha}\gamma_i(\alpha)C(\alpha),\quad\qquad\qquad N_i=\sum_{\beta}\sigma_i(\beta)N(\beta) \ .
\end{equation}
Here, the $\gamma_i(\alpha)$'s and $\sigma_i(\beta)$'s are the expansion coefficients, $C(\alpha)$ is the color basis whose elements consist of products of structure constants, and $N(\beta)$ is the kinematic basis, the elements of which are products of polarization vectors and momenta. 

The double copy procedure consists of exchanging the color factors $C_i$ in the numerator on the RHS of Eq. \eqref{A_YM} for a second instance of kinematic factors $\tilde N_i$, which in general may be taken from a different Yang-Mills theory and thus differ from the $N_i$'s. Remarkably, this replacement gives rise to a gravitational scattering amplitude,
\begin{equation}
A_\text{G}=\sum_i\frac{\tilde N_i\,N_i}{D_i} \equiv \sum_{\beta}\tilde N(\beta)A_\text{YM}(\beta) \equiv\sum_{\alpha \beta} N(\alpha) \tilde N(\beta)A_\text{S} (\alpha|\beta) \ , \label{ampl}
\end{equation}
where $A_\text{YM}(\beta)$ is the \emph{color-ordered} Yang-Mills amplitude given by
\begin{equation}
A_\text{YM}(\beta)=\sum_{\alpha} N(\alpha)A_\text{S}(\alpha|\beta) \ ,
\end{equation}
and $A (\alpha|\beta)$ is the \emph{doubly color-ordered} bi-adjoint scalar amplitude, 
\begin{equation}
A_\text{S} (\alpha|\beta)=\sum_i\frac{\gamma_i(\alpha)\sigma_i(\beta)}{D_i} \ .\label{ampl3}
\end{equation}
Different choices of kinematic factors $N_i$ and $\tilde N_i$ yield gravitational amplitudes with the same number of external gravitons but different intermediate states. As we will see in the next section, the Kerr-Schild formulation of the classical double copy will be somewhat reminiscent of the relations \eqref{ampl}--\eqref{ampl3}, although to the best of our knowledge the exact connection remains to be worked out.  

It is also worth mentioning that, besides the double copy procedure, other relations between scattering amplitudes have been shown to exist---see for instance~\cite{Cheung:2017ems}. One such relation corresponds to the multiple trace operation of \cite{Cheung:2017ems}, which relates a gravity amplitude to an Einstein-Maxwell one. This operation consists of applying trace operators $\tau_{ij}$ to the original amplitude. The trace operator is defined as $\tau_{ij}=\partial_{e_i\cdot e_j}$, where $e_i$ denotes the polarization vector of the particle $i$. Each trace operator reduces the spin of particles $i$ and $j$ by 1, and places them in a color trace. Applying these trace operators to a graviton amplitude exchanges some of the external gravitons for photons, which leads to an Einstein-Maxwell amplitude. In Sec. \ref{EM}, we will suggest a classical counterpart of this relation. A similar relation exists between pure Yang-Mills and Yang-Mills-scalar amplitudes, where the Yang-Mills and scalar field are coupled with the usual gauge interactions~\cite{Cheung:2017ems}.

\section{The classical double copy} \label{sec:classical double copy}

Let us now turn our attention to the classical double copy, first introduced in \cite{Monteiro:2014cda}. In its simplest implementation, one considers a space-time with a metric that admits a Kerr-Schild form with a Minkowski base metric, i.e.
\begin{equation}
g_{\mu\nu}  = \eta_{\mu\nu} + \phi \, k_\mu k_\nu \ , \label{KSm}
\end{equation}
where $\phi$ is a scalar field, and $k_\mu$ is a vector that is null and geodetic with respect to both the Minkowski and the $g_{\mu\nu}$ metrics:
\begin{equation}
g^{\mu\nu}k_\mu k_\nu=\eta^{\mu\nu}k_\mu k_\nu=0, \qquad\qquad\quad k^\mu\nabla_\mu k^\nu=k^\mu\partial_\mu k^\nu=0 \ .
\end{equation}
For our purposes, the crucial property of a metric in Kerr-Schild form is that the Ricci tensor ${R^\mu}_\nu$ turns out to be linear in $\phi$ provided all indices are raised using the Minkowski metric~\cite{Stephani:2003tm}.

Starting from the metric \eqref{KSm} in Kerr-Schild form, one can define a ``single copy'' Yang-Mills field via
\begin{equation}
A_\mu^a=c^a k_\mu\phi \ , \label{A} 
\end{equation}
where the $c^a$ are constant but otherwise arbitrary color factors. Then, if $g_{\mu\nu}$ is a solution to Einstein's equations, the Yang-Mills field \eqref{A} is guaranteed\footnote{This is true as long as we pick the correct splitting between the null KS vector and the KS scalar. We will  discuss this further below.} to satisfy the Yang-Mills equations, provided the gravitational coupling is replaced by the Yang-Mills one,~i.e.
\begin{equation}
8\pi G\rightarrow g, \label{subs}
\end{equation}
and any gravitational source is replaced by a color source. In fact, because of the factorized nature of the ansatz \eqref{A}, this implies that the field $A_\mu \equiv k_\mu\phi$ defined without color factors satisfies Maxwell's equations, in which case the color charges can be thought of just as electric charges.\footnote{Magnetic charges are instead related to NUT charges~\cite{Luna:2015paa}, which we will not consider in this paper.} In what follows we will restrict our attention to $A_\mu$, which we will also refer to as a single copy, with a slight abuse of terminology.

We can further combine the Kerr-Schild scalar $\phi$ with two copies of the color factors to define a bi-adjoint scalar 
\begin{equation}
\phi^{a\, b}= c^a c'^{b}\phi \ , \label{phi}
\end{equation}
which satisfies\footnote{Again, this is true as long as we pick the correct splitting between the KS vector and the KS scalar.} the linearized equations $\bar \nabla^2 \phi^{a\, b} = c^a c'^{b} \bar \nabla^2 \phi = 0$.  As in the case of the gauge field, in the following we will restrict out attention to the field $\phi$ stripped of its color indices. It is worth emphasizing that the equations of motion for the single copy $A_\mu$ and the zeroth copy $\phi$ turn out to be linear precisely because of the Kerr-Schild ansatz~\cite{Monteiro:2014cda}.

It is interesting to notice that the expressions for the ``metric perturbation'' $k_\mu k_\nu \phi$, the single copy $A_\mu = k_\mu \phi$, and the zeroth copy $\phi$ bear a superficial and yet striking similarity with the BCJ amplitudes in Eqs. \eqref{ampl}--\eqref{ampl3}. Specifically, a comparison between the two double copy procedures would seem to suggest that the vector $k_\mu$ somehow corresponds to the kinematic factors $N(\alpha)$, while the scalar $\phi$ is the analogue of $A_S(\alpha|\beta)$. Finally, the color factors $c^a$ can be thought of as the analogue of the color factors $C(\alpha)$. Although an exact mapping between the two double copies has not yet been derived, several analyses suggest that they are indeed related~\cite{Monteiro:2014cda,Luna:2015paa,Ridgway:2015fdl,Neill:2013wsa,Luna:2016hge,Luna:2016due,Goldberger:2016iau,Goldberger:2017frp}.

\section{Extending the classical double copy to curved spacetime}

In the following sections, we extend the classical double copy procedure to curved, maximally symmetric spacetimes: AdS and dS. One example of the classical double copy in a maximally symmetric spacetime was already considered in~\cite{Luna:2015paa}, which studied the Taub-NUT-Kerr-de Sitter solution. Our goal here is to consider the double copy procedure in (A)dS more systematically, and to obtain a more complete understanding of what happens in curved backgrounds by finding additional examples in which the double copy procedure is applicable. 

To this end, we will use the generalized Kerr-Schild form of the metric
\begin{equation} \label{KS g bar}
g_{\mu\nu}=\bar g_{\mu\nu}+ \phi \, k_\mu k_\nu  \ ,
\end{equation} 
where the base metric $\bar g_{\mu\nu}$ is now (A)dS (unless otherwise specified), while $k_\mu$ is again null and geodetic with respect to both the full and base metrics. A detailed analysis of these kinds of metrics can be found in \cite{TAUB1981326,0264-9381-4-5-005,Malek:2010mh}. Even with a more general choice for the base metric, the Ricci tensor $R^{\mu}_{\ \nu}$ is still linear in $\phi$~\cite{Stephani:2003tm}: 
\begin{equation} \label{R curved}
R^\mu{}_\nu = \bar R^\mu{}_\nu -\phi k^\mu k^\lambda \bar R_{\lambda \nu} +\tfrac{1}{2} \left[ \bar \nabla^\lambda \bar \nabla^\mu (\phi k_\lambda k_\nu ) +   \bar \nabla^\lambda \bar \nabla_\nu (\phi k^\mu k_\lambda ) -  \bar \nabla^2 (\phi k^\mu k_\nu )  \right] \ ,
\end{equation}
and the single and zeroth copy of the metric \eqref{KS g bar} are still defined by Eqs. (\ref{A}) and (\ref{phi}) respectively.

At this point, we should discuss one aspect of the classical double copy construction that so far has not been mentioned in the literature but is nevertheless crucial to ensure that the classical double copy procedure gives rise to sensible results. For any given choice of coordinates that allows the metric to be written in the Kerr-Schild form, the null vector $k_\mu$ and the scalar $\phi$ are not uniquely determined, since Eq. \eqref{KS g bar} is invariant under the rescalings
\begin{equation}
k_\mu \to f k_\mu, \qquad \qquad \quad \phi \to \phi / f^2 \ , \label{resc}
\end{equation}  
for any arbitrary function $f$. If we demand that the null vector $k_\mu$ is geodetic, this imposes restrictions on $f$, but does not fix it completely. Of course, this ambiguity is immaterial when it comes to the gravitational theory, since the Ricci tensor in Eq. \eqref{R curved} is also invariant under this redefinition. However, the single and zeroth copies defined in Eqs. \eqref{A} and \eqref{phi} are not, and neither are the equations that they satisfy. It is worth stressing that this ambiguity is not a peculiarity of curved space-time, but a general feature of any metric in Kerr-Schild form.

To further illustrate this point, it is convenient to recast Eq. \eqref{R curved} in the following form:
\begin{equation}
2(\bar R^\mu{}_\nu - R^\mu{}_\nu)= \left[ \bar \nabla_\lambda F^{\lambda\mu} + \tfrac{(d-2)}{d(d-1)} \bar R A^\mu \right] k_\nu  + X^\mu{}_\nu +Y^\mu{}_\nu \ , \label{rr}
\end{equation}
where here and in what follows, $F^{\lambda\mu}$ is the usual field strength for an abelian gauge field\footnote{Our anti-symmetrization conventions are such that $B_{[\mu\nu]} \equiv B_{\mu\nu}-B_{\nu\mu}$.}, and its components have been raised using the base metric. Moreover, we have simplified our notation by introducing the following quantities:

\begin{align}
X^\mu{}_\nu\equiv& - \bar\nabla_\nu\left[A^\mu\left(\bar\nabla_\lambda k^\lambda+\frac{k^\lambda \bar\nabla_\lambda \phi}{\phi}\right)\right] \ ,\\
Y^\mu{}_\nu\equiv& \, F^{\rho\mu}\bar{\nabla}_\rho k_\nu-\bar{\nabla}_\rho\left(A^\rho \bar{\nabla}^\mu k_\nu-A^\mu \bar{\nabla}_\rho k_\nu \right) \ .
\end{align} 
When the full metric solves the Einstein equations with a cosmological constant, the LHS is equal to $-16\pi G\left({T^\mu}_\nu-\delta^\mu_\nu T/(d-2)\right)$, with ${T^\mu}_\nu$ the stress-energy tensor. If we contract Eq. \eqref{rr} with a Killing vector $V^\nu$ of both the base and full metric, we obtain the equation of motion for the single copy $A^\nu$ in $d$ dimensions: 
\begin{equation}
 {\bar\nabla}_\lambda F^{\lambda\mu} + \tfrac{(d-2)}{d(d-1)} \bar R A^\mu  + \tfrac{V^\nu}{V^\lambda k_\lambda}\left(X^\mu{}_\nu + Y^\mu{}_\nu\right) =8\pi G\,  J^\mu \ , \label{Aeom}
\end{equation}
where we have defined
\begin{equation}
J^\mu\equiv-\tfrac{2 V^\nu}{V^\rho k_\rho}\left({T^\mu}_\nu-\delta^\mu_\nu \tfrac{T}{d-2}\right). \label{current}
\end{equation}
To obtain the zeroth copy equation, we can further contract Eq. \eqref{rr} with another Killing vector $V_\mu$ and find
\begin{equation}
\bar \nabla^2\phi=j-\tfrac{(d-2)}{d(d-1)} \bar R \phi -\tfrac{V_\nu}{( V^\mu k_\mu)^2}\left(V^\mu X^\nu{}_\mu + V^\mu Y^\nu{}_\mu+ Z^\nu\right) \label{phieom}
\end{equation}
where
\begin{equation}
Z^\nu\equiv( V^\rho k_\rho)\bar\nabla_\mu\left(\phi\bar\nabla^{[\mu}k^{\nu]}- k^\mu\bar\nabla_\nu\phi\right),
\end{equation}
and the source is defined as
\begin{equation}
j=\frac{V_\nu J^\nu}{V^\rho k_\rho}.
\end{equation}

In what follows, we will use the timelike Killing vector for stationary solutions, and  the null Killing vector for wave solutions. The Killing vector allows us to find the correct sources for the single and zeroth copies. Clearly, Eq. \eqref{Aeom} is not invariant under the rescaling of Eq. \eqref{resc}. This freedom allows us to choose the null vector and scalar such that the copies satisfy `reasonable' equations of motion. By this, we mean that when there is a localized source on the gravitational side, we obtain a localized source in the gauge and scalar theories; when there is no source for Einstein's equations, there is no source in the Abelian Yang-Mills and scalar equations either. At this stage, we are unable to formulate a more precise criterion that selects the correct splitting between $\phi$ and $k_\mu$ based on fundamental principles. However, we believe this is an important question to address in future work and we will touch upon it again in the final section of this paper.

Before constructing explicit examples, it is worth discussing a few details of the two cases of interest in this paper: stationary spaces and waves. First, note that the terms in $X^\mu{}_\nu$ in the inner parentheses correspond to the expansion of $k^\mu$ and the derivative of $\phi$ along the direction of $k^\mu$. For stationary solutions, where $\bar\nabla_\lambda k^\lambda\neq0$ and $k^\lambda \bar\nabla_\lambda \phi\neq0$ , $X^\mu{}_\nu$ is non-zero. In these cases, we may choose $k_\mu V^\mu$ (which corresponds to choosing the scaling function $f$ between $k_\mu$ and $\phi$) such that the single copy satisfies Maxwell's equations. On the other hand, for wave solutions the expansion term is zero and the null vector is orthogonal to the gradient of $\phi$, so that $X^\mu{}_\nu=0$. In order to obtain a reasonable equation of motion for the wave solutions, we require that the terms in $ Y^\mu{}_\nu$ that contain derivatives of the gauge field cancel out. These terms can be rewritten as $ Y^\mu{}_\nu\supset \left(F^{\rho\mu}+\nabla^\rho A^\mu\right) \nabla_\rho k_\nu$; setting them to zero is equivalent to choosing $k_\mu$ such that $\bar\nabla_\rho k_\nu=0$, with $\rho\neq u$, for a wave traveling in the direction of the light-cone coordinate $u$. One can see that this choice will in fact set $Y^\mu{}_\nu=0$. This choice doesn't completely fix $k_\mu$; in fact, we can still do a rescaling as in Eq.\eqref{resc} with $f=f(u)$. The fact that we can re-scale our solution in such a way is a property of wave solutions; multiplying by $f(u)$ only changes the wave profile. Remarkably, we seem unable to choose $V^\mu k_\mu$ in such a way as to also cancel the second term in Eq. \eqref{Aeom}.  Therefore, in this case the single copy satisfies an equation in which the gauge symmetry is broken by a non-minimal coupling to the background curvature. Furthermore, once we have fixed the splitting we find that
\begin{equation}
\frac{V^\nu Z^\nu}{V^\rho k^\rho} =-\frac{2(d-3)}{d(d-1)}\bar R\phi,
\end{equation} 
in both cases. This means that the zeroth copy has a mass proportional to the Ricci scalar. For the special case of solutions in 4d, the stationary solutions follow the conformally invariant equation and the wave solutions the equation for a massless scalar.

\section{(A)\lowercase{d}S-Schwarzschild} \label{Sch}

The simplest example of the classical double copy procedure in maximally symmetric curved spacetimes is the (A)\lowercase{d}S-Schwarzschild black hole in $d=4$ space-time dimensions. In order to find the corresponding single and zeroth copies, we write this solution in the Kerr-Schild form, using an (A)dS base metric in global static coordinates,
\begin{equation}
\bar g_{\mu\nu} \ud x^\mu \ud x^\nu = -\left(1-\frac{\Lambda \, r^2}{3} \right) \ud t^2 + \left(1-\frac{\Lambda \, r^2}{3} \right)^{-1} \ud r^2 + r^2 \ud \Omega^2 
\end{equation}
with $\Lambda$ the cosmological constant, and choose the null vector $k_\mu$ and scalar function $\phi$ in the following way:
\begin{equation}
k_\mu dx^\mu  = \ud t + \frac{\ud r}{1-\Lambda \, r^2/3}\ , \qquad \qquad\qquad \phi =\frac{2 G M}{r} \ , \label{kmu phi SAdS}
\end{equation}
In this case, the full metric $g_{\mu\nu}$ defined by \eqref{KS g bar} is a solution to the Einstein equations 
\begin{equation}
G_{\mu\nu}+\Lambda g_{\mu\nu}=8\pi GT_{\mu\nu} \ .
\end{equation}
One can remove the singularity at $r=0$ by including a localized source with stress-energy tensor
\begin{equation}
T^\mu{}_\nu =\frac{M}{2}\text{diag} (0,0,1,1) \delta^{(3)} (\vec{r}) \ , \label{TS}
\end{equation}
where the $\delta^{(3)} (\vec{r})$ should be expressed in spherical coordinates. Thus, the only two nonzero components of the stress-energy tensor in Kerr-Schild coordinates are the angular ones along the diagonal.

At first sight, this result might seem at odds with one's ``Newtonian intuition''. In fact, in the usual Newtonian limit the dominant component of Einstein's equation is the $(t,t)$ one, which reduces to a Poisson's equation for $\delta g_{tt}$ sourced by $T_{tt}$. However, this result is based on the assumption that the off-diagonal components of the metric are negligible. In Kerr-Schild coordinates, the $\delta g_{tr}$ components are of the same order in $GM/r$ as the perturbations to the diagonal components of the metric---they are all of order $\phi$. In this case, the Poisson's equation for the Newtonian potential arises from the angular components of Einstein's tensor, rather than from the $(t,t)$ component---a result in agreement with what was found in Sec. III of~\cite{Ridgway:2015fdl}.

We can now follow the procedure discussed in Sec. \ref{sec:classical double copy} to construct the single and zeroth copies. It is in fact easy to show using the replacement rule \eqref{subs} that the single copy $A_\mu=k_\mu\phi$ with $k_\mu$ and $\phi$ given by Eq. \eqref{kmu phi SAdS} satisfies Maxwell's equations on an (A)dS background,
\begin{equation}
\bar \nabla_\mu F^{\mu\nu}=g\,J^\nu \ , \label{YMeom} 
\end{equation}
with a localized, static source given by 
\begin{equation}
J^\mu = M\, \delta^{(3)}(\vec{r}) \, \delta^\mu_0 \ .
\end{equation}
This source can be derived from Eq. \eqref{current} using the timelike Killing vector of the Schwarzschild metric.  As expected, this source describes a static point-particle with charge $Q=M$ in (A)dS, in perfect analogy with the flat case. 

The zeroth copy $\phi$ instead satisfies the equation of motion
\begin{equation}
\left(\bar \nabla^2-\frac{\bar R}{6}\right)\phi= j \ ,  \label{seom}
\end{equation}
with a localized source $j = M\, \delta^{(3)}(\vec{r})$. Thus, moving away from a flat background it becomes apparent that the zeroth copy satisfies the equation for a conformally coupled scalar field rather than simply $\bar \nabla^2 \phi = j$. This was first noticed in \cite{Luna:2015paa}, where it was also argued that this might be tied to the conformal symmetry of the Yang-Mills equations in $d=4$. In fact, in $d \neq 4$ the non-minimal coupling between $\phi$ and the Ricci scalar does not have a conformal value~\cite{Luna:2015paa}, as we will see in the next section.

Let us now restrict our attention to the dS solution and consider the case of small dS black holes, i.e. black holes such that $0<M<M_\text{max}\equiv1/(3G\Lambda^{1/2})$. This spacetime has both a cosmological horizon and a black hole horizon. As the mass increases, the black hole horizon grows and the cosmological horizon shrinks. At the particular value $M=M_\text{max}$, both horizons have the same area but the distance between them remains finite. In this limit, the singularity disappears and the patch between the two horizons corresponds to dS$_2\times S^2$. This spacetime is known as Nariai solution~\cite{nariai1950some,nariai1951some}. At the level of the single and zeroth copies, however, there is no privileged value that the charge can take---all solutions to Eqs. \eqref{YMeom} and \eqref{seom} look qualitatively the same regardless of the value of $M$. This can be easily understood from the fact that, if there existed a special value for the charge $Q_*$, then in the spirit of the double copy procedure it should be such that $Q_* \propto M_{\rm max} / M_{\rm pl} \propto 1/\sqrt{G \Lambda}$. However, by considering a fixed background for the single and zeroth copy, we are effectively working in the limit $G \Lambda \to 0$ while keeping $M^2 /\Lambda$ constant, in which case $Q_* \to \infty$. Another point of view is that on the gravitational side, $M=M_{\rm max}$ is a special value due to the existence of horizons. This is strictly a gravitational property which does not have an analogue in Yang-Mills or scalar theories.

Finally, it is instructive to discuss how the wrong choice of the Kerr-Schild vector $k_\mu$ can give rise to an unreasonable double copy. Instead of the definitions in Eq.\eqref{kmu phi SAdS}, we will pick the splitting such that
\begin{equation}
k_\mu dx^\mu  = f(\theta) \left( \ud t + \frac{\ud r}{1-\Lambda \, r^2/3} \right) \ , \qquad\qquad \phi =\frac{1}{f(\theta)^2}\frac{2 G M}{r} \ , \label{Alt kmu phi SAdS}
\end{equation}
where $f(\theta)$ is an arbitrary function. This choice preserves several properties of the Kerr-Schild vector, namely $k_\mu$ is null, geodetic, shear-free, and twist-free\footnote{In fact, $k_\mu$ is null, geodetic, shear-free, and twist-free for an arbitrary function $f(\theta,\phi)$, but we restrict ourselves to $f(\theta)$ for simplicity.}. As before, we can define the single copy as $A_\mu=k_\mu\phi$ and find that it satisfies Eq.\eqref{YMeom} with a source current given by
\begin{equation}
J^\mu = M\, \delta^{(3)}(\vec{r}) \, \delta^\mu_0 + \tilde{j}^\mu \ ,
\end{equation}
where
\begin{equation}
\tilde{j}^\mu=-\frac{g(\theta)}{r^2
	f(\theta)^3 }\left(\frac{\delta^\mu_t}{\left(1-\Lambda  r^2/3\right)}+ \delta^\mu_r\right) , \quad g(\theta)=2 f'(\theta)^2-f(\theta) \left(f''(\theta)+\cot
	(\theta) f'(\theta)\right) \ .
\end{equation}
This extra term in the current is clearly non-localized and changes the total charge. Given our criteria for a reasonable single copy, this term is unacceptable. We conclude that the choice Eq.\eqref{Alt kmu phi SAdS} with an arbitrary function $f(\theta)$ is incorrect. We can see that, taking $f(\theta)=1$ sets $\tilde{j}=0$ recovering the correct result for the single copy.

\section{Kerr-(A)\lowercase{d}S} \label{kerr}

We now consider a more involved example, namely that of a rotating black hole in (A)dS. As in the previous section, we will derive the single and zeroth copies of the Kerr-(A)dS solution in $d=4$ by casting the full metric in a Kerr-Schild form. To this end, it is convenient to express the base (A)dS metric in spheroidal coordinates~\cite{Gibbons:2004uw},
\begin{align}
\bar g_{\mu\nu} \ud x^\mu \ud x^\nu = -\frac{\Delta}{\Omega}\, f\,  \ud t^2+ \frac{\rho^2 \ud r^2}{(r^2+a^2)f} + \frac{\rho^2 \ud \theta^2}{\Delta} + \frac{(r^2+a^2)}{\Omega}\sin^2 \theta \, \ud \varphi^2 \ ,
\end{align}
where we have defined
\begin{equation}
f \equiv 1-\frac{\Lambda \, r^2}{3}, \qquad \rho^2\equiv r^2+a^2\cos^2{\theta},\qquad \Delta\equiv 1+\frac{\Lambda}{3}a^2\cos^2{\theta},\qquad \Omega\equiv 1+\frac{\Lambda}{3}a^2 \ .
\end{equation}
The corresponding null vector and scalar function read~\cite{Gibbons:2004uw}
\begin{align}
k_\mu \ud x^\mu = \frac{\Delta \ud t}{\Omega} + \frac{\rho^2 \ud r}{(r^2+a^2)f} - \frac{a\sin^2{\theta} \, \ud \theta}{\Omega} \ , \qquad \qquad \phi =\frac{2 M G\, r}{\rho^2}.
\end{align}
It is easy to see that, when $a \to 0$, these expressions reduce to the ones used in the previous section for the (A)dS-Schwarzschild solution.  Notice also that the Kerr-(A)dS black hole solution becomes singular if $|a|\geq\sqrt{3 /(-)\Lambda}$.

The source of the metric corresponds to a negative proper surface density given by a disk of radius $a$ localized at $r=0$.  This can be seen by considering the induced metric at $r=0$ which corresponds to the history of the disk,
\begin{equation}
\ud s^2=-\Delta \ud \tilde t^2 + \Delta^{-1} \ud R^2 + R^2  \ud \tilde \varphi^2,
\end{equation} 
where the new coordinates are given by
\begin{equation}
\tilde{t}=t/\Omega,\qquad\qquad R=a \sin{\theta},\qquad\qquad\tilde{\varphi}=\varphi/\Omega.
\end{equation} 
This disk is rotating about the $z$ axis with superluminal velocity and is balanced by a radial pressure. The corresponding stress-energy tensor can be written as
\begin{equation}
T^{\mu}{}_{\nu}=-j \cos^2{\theta}\left(\xi^\mu\xi_\nu+u^\mu u_\nu\right),\qquad\qquad j=\frac{M}{4 \pi \,a^2}\sec^3{\theta}\, \delta(r) \ ,  \label{jkerr}
\end{equation}
where
\begin{equation}
\xi_\mu= \frac{a\cos{\theta}}{\Delta^{1/2}}\, (0,0,1,0),\qquad\qquad  u_\mu= \frac{\Delta^{1/2} \tan{\theta}}{\Omega} \, \left(1,0,0,-a\right)
\end{equation}
This is the (A)dS generalization of the source for the flat Kerr solution given in \cite{Israel:1970kp}.

The single copy solution is given as usual by $A_\mu=k_\mu\phi$, with the substitution~\eqref{subs}, and it again satisfies the Maxwell equation~\eqref{YMeom}, with the source now given by
\begin{equation}
J^\mu=j\, \zeta^\mu, \qquad\qquad \zeta_\mu=(2,0,0,2/a) \ .
\end{equation}
As expected, the single copy corresponds to the field generated by a charged disk rotating around the $z$ direction in (A)dS spacetime. This field generates both an electric and a magnetic field, with the latter proportional to the angular momentum of the charged particle. Thus, the angular momentum on the gravity side is translated into a magnetic field at the level of the single copy. As we will see in Sec.~\ref{btz}, the same correspondence will hold also for the BTZ black hole.  In a similar way, the scalar field satisfies Eq.~\eqref{seom} with source $2j$, where $j$ is given as in Eq.~\eqref{jkerr}. 

The previous analysis can easily be extended to higher dimensions. In fact, the Myers-Perry black hole with a non-vanishing cosmological constant also admits a Kerr-Schild form~\cite{Gibbons:2004uw}. In $d=2n+1$, the null vector and scalar field read
\begin{align}
k_\mu \ud x^\mu =W \ud t + F \ud r -\sum_{i=1}^{n}\frac{a_i\mu_i^2}{1+\lambda a_i^2}\ud\varphi_i \ ,\qquad 
\phi =\frac{2GM}{\sum_{i=1}^{n}\frac{\mu_i^2}{r^2+a_i^2}\prod_{j=1}^{n}(r^2+a_j^2)} \ ,
\end{align}
where $\lambda = \frac{2 \Lambda}{(d-2)(d-1)}$, 
\begin{equation}
W\equiv\sum_{i=1}^{n}\frac{\mu_i^2}{1+\lambda a_i^2} \ ,\qquad\qquad F\equiv\frac{r^2}{1- \lambda r^2}\sum_{i=1}^{n}\frac{\mu_i^2}{r^2+a_i^2} \ ,
\end{equation}
and the coordinates $\mu_i$ are subject to the constraint
\begin{equation}
\sum_{i=1}^{[d/2]} \mu_i=1 \ .
\end{equation}
Meanwhile, for $d=2n$ we instead have 
\begin{align}
k_\mu \ud x^\mu =W \ud t + F \ud r -\sum_{i=1}^{n-1}\frac{a_i\mu_i^2}{1+\lambda a_i^2}\ud\varphi_i \ , \qquad \phi =\frac{2GM}{r \sum_{i=1}^{n}\frac{\mu_i^2}{r^2+a_i^2}\prod_{j=1}^{n-1}(r^2+a_j^2)} \ .
\end{align}
When all the rotation parameters $a_i$ vanish, the metrics above reduce to the higher-dimensional version of the (A)dS-Schwarzschild one, and the corresponding source becomes a static charge in (A)dS$_d$.

Constructing the single and zeroth copy is very similar to the $d=4$ case. In particular, the corresponding gauge field is sourced by a charge rotating with angular momentum proportional to $a_i$ in the corresponding directions. Most interestingly, the zeroth copy satisfies the equation
\begin{equation}
\left(\bar \nabla^2- \frac{2(d-3)}{d(d-1)} \bar R\right)\phi=j \ , \label{scalar} 
\end{equation}
where the non-minimal coupling to the curvature has a conformal value only in $d=4$. We have explicitly checked the coefficient of the Ricci scalar for $4 \leq d \leq 11$. The result \eqref{scalar} remains valid in the limit $a \to 0$, and as such it generalizes Eq. \eqref{seom} for a (A)dS-Schwarzschild black hole to arbitrary dimensions.\footnote{The equation for the zeroth copy of dS-Schwarzschild black holes in arbitrary dimensions already appeared in a presentation given by Andres Luna at the conference ``QCD meets Gravity'', UCLA, Dec 2016~\cite{UCLA-link}.}

\section{Charged black hole solutions} \label{EM}

It is interesting to observe that the single copy we built for the (A)dS-Schwarzschild and (A)dS-Kerr black holes is automatically a solution to the Einstein-Maxwell's equation when the metric is promoted to its charged version~\cite{Romans:1991nq}, i.e. A(dS)-Reissner-Nordstrom and A(dS)-Kerr-Newman respectively. Moreover, there is a simple procedure that yields these charged metrics starting from their neutral counterparts in Kerr-Schild coordinates.

To illustrate this, consider a neutral black hole metric in Kerr-Schild form 
\begin{equation}
g_{\mu\nu}= \bar g_{\mu\nu} + k_\mu k_\nu \phi(8 \pi G M) \ ,
\end{equation}
where for later convenience we have explicitly shown the dependence of $\phi$ on the gravitational coupling and the black hole mass. As shown in previous examples, one can define the corresponding single copy as $A_\mu=k_\mu\phi(g Q)$, where we have made the substitution $M \to Q$ and have applied the replacement rule \eqref{subs}. Of course, in the case of an Abelian gauge theory the coupling $g$ in \eqref{subs} is redundant, and in what follows we will set $g=1$ by an appropriate rescaling of the charge $Q$. 

Before constructing the charged black hole solutions, we analyze Kerr-Schild solutions in Einstein-Maxwell theory; for a review of these types of solutions see \cite{Griffiths:2009dfa,Stephani:2003tm}. The fact that the metric is in Kerr-Schild form imposes restrictions on the stress-energy tensor that can be translated into restrictions on the field strength $F^{\mu\nu}$ when the matter is a $U(1)$ field. When the null KS vector is geodetic and shear-free, it should also be an eigenvector of the Maxwell field strength
\begin{equation}
k_\mu F^{\mu\nu}=\lambda k^\nu \ .
\end{equation}
This requirement is a necessary but not sufficient condition for the gauge field to be a solution of the field equations. As a consistency check, we confirm that the single copy ansatz $A^{EM}_\mu=k_\mu \phi$ satisfies the above requirement with an eigenvalue
\begin{equation}
\lambda=k_\mu\nabla^\mu\phi \ .
\end{equation}

We now show how to construct charged black holes using the single copy. Using the Kerr-Schild ``building blocks'' above, we can immediately write down an electrically charged solution to the Einstein-Maxwell equations, where the metric and gauge field are given by
\begin{align}
g_{\mu\nu}^\text{EM} = \bar g_{\mu\nu} + k_\mu k_\nu\, \phi^\text{EM}(M,Q) \ , \qquad \qquad A^{EM}_\mu = A_\mu = k_\mu\phi(Q) \ ,
\end{align}
with
\begin{equation}
\phi^\text{EM}(M,Q)=\phi(8 \pi G M)- \frac{Q}{r^{d-3}} \phi(Q)\ .
\end{equation}
This construction works both in curved and flat space, but it is not applicable in $d<4$ because in that case there are no graviton degrees of freedom. 

This ``recipe'' allows us to turn a solution to Einstein's equations into one that satisfies the Einstein-Maxwell equations. This is somewhat reminiscent of the transmutation operations for scattering amplitudes described in \cite{Cheung:2017ems}. In particular, the procedure we have described appears to be a classical analog of the multiple trace operation that turns gravity amplitude into Einstein-Maxwell ones. However, significantly more evidence is required to establish if there is a connection between these two procedures.

\section{Black strings and black branes} \label{others}

Black strings and black branes are black hole solutions with extended event horizons. In this section, we construct their corresponding single and zeroth copies in (A)dS in $d>4$ spacetime dimensions.

\subsection{Black strings}

In order to construct black strings in AdS$_d$, we start from the base metric  
\begin{equation}
\bar g_{\mu\nu} \ud x^\mu \ud x^\nu = a^2(z)\left(\bar \gamma_{ab}\ud x^a \ud x^b+\ud z^2\right) \ ,
\end{equation}
where the $(d-1)$-dimensional metric $\bar \gamma_{ab}$ can be that of (A)dS$_{d-1}$ or Mink$_{d-1}$, depending on whether one chooses a (A)dS or a Minkowski slicing of AdS$_d$. The corresponding form of the scale factor is~\cite{Hirayama:2001bi}
\begin{align}
a^{-1}(z)=\begin{cases}
\ell_{d-1} / \ell_d\,\sinh{\left(z/\ell_{d-1}\right)} & \qquad\quad  \text{(dS slicing)}\\
z / \ell_d & \qquad\quad  \text{(Minkowski slicing)}\\
\ell_{d-1} / \ell_d\,\sin{\left(z/\ell_{d-1}\right)} & \qquad\quad  \text{(AdS slicing)}
\end{cases} \ ,
\end{align} 
where $\ell_d$ and $\ell_{d-1}$ are the AdS length scales in $d$ and $d-1$ dimensions, respectively. A black string solution is then obtained by replacing the $(d-1)$-dimensional metric $\bar \gamma_{ab}$ with a $(d-1)$-dimensional Schwarzschild black hole with the same cosmological constant~ \cite{Chamblin:1999by,Gregory:2000gf}.
 
A similar construction is possible for a de Sitter black string starting from a dS$_d$ space foliated by dS$_{d-1}$, in which case the base metric reads
\begin{equation}
	\bar g_{\mu\nu} \ud x^\mu \ud x^\nu = \sin^2(z/\ell_d) \, ds^2_{{\rm dS},d-1} + dz^2  \ .
\end{equation}
The metric for a black string in dS is then obtained by replacing $ds^2_{{\rm dS},d-1}$ with the line element for a dS-Schwarzschild black hole in $(d-1)$-dimensions.

In both the AdS and dS cases, if the black hole metric is in the Kerr-Schild form, the full metric automatically inherits a similar form. More precisely, writing the black hole metric as
\begin{equation}
\gamma_{\mu\nu}=\bar{\gamma}_{ab}+ \psi  k_a k_b \ ,
\end{equation}
the null vector $k_\mu$ and scalar $\phi$ for the (A)dS black string can be chosen to be
\begin{equation}
k_\mu= a^{2}(z) \delta_\mu^a k_a \ , \qquad \qquad \quad \phi = \frac{\psi}{a^2(z)} \ . \label{k phi black strings}
\end{equation}
The stress-energy tensor for the AdS black string in Kerr-Schild coordinates reads
\begin{equation}
T^\mu{}_\nu =\frac{m}{2 a(z)^{2} }\, \text{diag} (0,0,1,1, \vec 0) \, \delta^{(d-2)} (\vec{r}) \ ,\label{TString}
\end{equation}
where $m$ is the mass per unit length of the string. 

It is now easy to show that the single copy $A_\mu = \phi \, k_\mu$ and the zeroth copy $\phi$ satisfy the equations  (\ref{YMeom}) and (\ref{seom}) respectively with sources:
\begin{equation}
J^\mu=j \, \delta^\mu_0 \ ,\qquad\qquad\quad  j=\frac{m}{a^4(z)} \, \delta^{(d-2)} (\vec{r}) \ .
\end{equation} 
As expected, the YM source is a charged line aligned with the $z$ direction with charge per unit length $q = m$, living in either AdS or dS.  Notice that the judicious insertion of scale factors in Eq. \eqref{k phi black strings} is crucial to obtaining sensible classical copies.

\subsection{Black branes}

We now turn to the case of black branes, or planar black holes, in AdS$_{p+2}$ . The most familiar form of the metric for black branes is
\begin{equation}
\ud s^2=\frac{r^2}{\ell^2}\left\{-\left[1-\left(\frac{r_h}{r}\right)^{p+1}\right]\ud t^2+\eta_{ab} \ud x^a \ud x^b \right\} + \frac{\ell^2}{r^2} \left[1-\left(\frac{r_h}{r}\right)^{p+1}\right]^{-1}\ud r^2  \ , \label{metric brane}
\end{equation}
where $a,b = 1,\dots,p$, and the horizon is located at $r=r_h$. This is a solution to Einstein's equations with a source
\begin{equation}
	T^\mu{}_\nu = \frac{r_h^{p+1}}{2 \ell^2} \,  \text{diag} (0,0,1,1, ... , 1) \, \delta(r) \ .
\end{equation}
The metric in Eq. \eqref{metric brane} can be put in a Kerr-Schild form by introducing a new time coordinate. The base metric is then just AdS in Poincar\'e coordinates, and  the Kerr-Schild null vector and scalar can be chosen as follows:
\begin{align}
k_\mu \ud x^\mu & = d\tau - \frac{\ell^2}{r^2} dr \ , \qquad \qquad \quad \phi= \frac{r^2}{\ell^2} \left(\frac{r_h}{r}\right)^{p+1} \ .
\end{align}
The single copy given by $A_\mu=k_\mu\phi$ satisfies the  Abelian Yang-Mills equations of motion with a source 
\begin{equation}
	J^\mu =  j  \delta^\mu_0 \delta(r) \ , \qquad \qquad j = \frac{r_h^{p+1}}{\ell^2} \delta (r) \ .
\end{equation} 
which gives rise to an electric field in the $r$ direction. Meanwhile, the scalar field satisfies Eq. \eqref{scalar} with source $j$.

\section{Wave solutions} \label {waves}

We now turn our attention to time-dependent solutions, and in particular to wave solutions. For negative values of the cosmological constant, there are three different types of wave solutions in vacuum that can be written in the Kerr-Schild form: {\it Kundt waves}, {\it generalized pp-waves}, and {\it Siklos waves} \cite{Griffiths:2003bk,Griffiths:2009dfa}. All of these solutions are Kundt spacetimes of Petrov-type N. By contrast, in the case of a positive cosmological constant, there is only one kind of wave in vacuum --- Kundt waves --- which in this case are the same as pp-waves~\cite{Griffiths:2009dfa}. Finally, we consider \emph{shock waves}, which unlike the previous solutions are generated by a non-trivial localized source.  Since these spacetimes are not stationary, there is no timelike Killing vector. However, all these cases feature a null Killing vector, which we can use to construct the classical single and zeroth copies.  As in previous cases, the ambiguity in choosing the form of the null KS vector and the KS scalar will play a crucial role in ensuring the existence of reasonable single and zeroth copies. Unlike the stationary cases, here we have the freedom of performing a rescaling as in Eq.\eqref{resc} with $f=f(u)$; such rescaling is a property of wave solutions and it only changes the wave profile. We will find that the single and zeroth copy satisfy the same equations in all of these cases (albeit with a source term in the case of shock waves). However, the equation for the single copy is no longer gauge invariant when the base metric is curved. For simplicity, in this section we will restrict ourselves to $d=4$ spacetime dimensions.

\subsection{Kundt waves}

We begin by analyzing the case of Kundt waves, which exist in both de Sitter and anti-de Sitter spacetimes. The Kundt waves in (A)dS can be written in Kerr-Schild form with a base metric that reads
\begin{equation}
\bar g_{\mu\nu} \ud x^\mu \ud x^\nu  = \frac{1}{P^2}\left[-4 x^2 \ud u\left(\ud v - v^2 \ud u\right)+\ud x^2 + \ud y^2\right] \ ,\qquad P=1+\frac{\Lambda}{12}(x^2+y^2) \ , \label{wdS}
\end{equation}
where $u$ and $v$ are light-cone coordinates. The null vector and the scalar are given by
\begin{align}
k_\mu =\frac{x}{P} \, \delta_\mu^u \ ,\qquad \qquad \quad \phi= \frac{P}{x} H(u,x,y) \ .
\end{align}
The full metric $g_{\mu\nu} = \bar g_{\mu\nu} + \phi \, k_\mu k_\nu$ is a vacuum solution to the Einstein equations provided $H(u,x,y)$ satisfies the following partial differential equation:
\begin{equation}
\left[ \partial_x^2 + \partial_y^2 + \frac{2\Lambda}{3 P^2} \right] H(u,x,y) = 0 \ . \label{eq H Kundt}
\end{equation}

\begin{figure}[!t] 
	\includegraphics[scale=0.5]{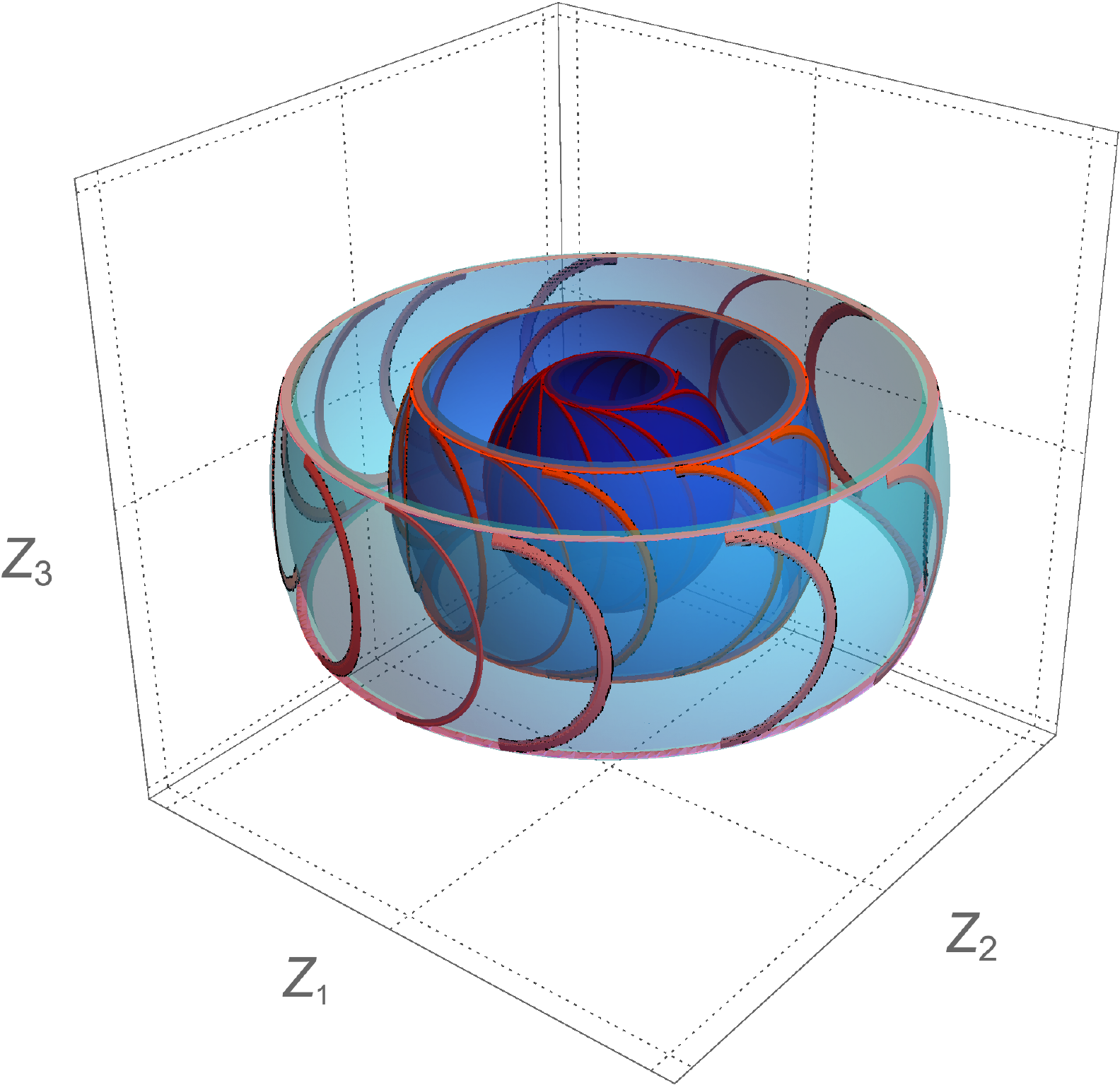}
	\caption{Kundt waves in de Sitter space. We can consider dS as a four-dimensional hyperboloid embedded in a flat 5d space with coordinates $Z^a$. This figure shows portions of de Sitter covered by the coordinates in Eq.~\eqref{wdS} at different values of the time $Z^0$ with $Z^4=0$. The portions of 2-spheres correspond to different snapshots in time and the semi-circles on them are the wavefronts of constant $u$. The gravitational, gauge, and scalar waves all have wavefronts of this shape. For more details, see \cite{Griffiths:2009dfa,Griffiths:2003bk}.}
	\label{dswave}
\end{figure}

The singularity of the metric Eq.~\eqref{wdS} at $x=0$ corresponds to an expanding torus in de Sitter, and to an expanding hyperboloid in anti-de Sitter. In dS, the wavefronts are tangent to the expanding torus and correspond to hemispheres with constant area $4 \pi \ell^2$, with $\ell = \sqrt{3/\Lambda}$ the dS radius---see Fig.~\ref{dswave}. For AdS, the wave surfaces are semi-infinite hyperboloids. In both cases, the wavefronts are restricted to $x\geq 0$ to avoid caustics (except for the singularity $x=0$)~\cite{Griffiths:2003bk} and different wave surfaces are rotated relative to each other. It should also be noted that the wave surfaces in the dS and AdS cases only exist outside the expanding singular torus or hyperboloid respectively. 

Contrary to what we have seen in the time-independent cases, in this case the gauge field $A_\mu = \phi \, k_\mu$ and scalar field $\phi$ satisfy the following equations:
\begin{eqnarray}
&\displaystyle \bar \nabla_\mu F^{\mu\nu}+\frac{\bar R}{6}A^\nu=0 \ , & \label{AeqW} \\
&\bar \nabla^2\phi=0 \ . & \label{SeqW}
\end{eqnarray}
This can be seen by using the $(\mu,u)$ component of Einstein's equations, and the equation for $H(u,x,y)$ in \eqref{eq H Kundt}. The copies correspond to waves in the gauge and scalar theory whose wavefronts are the same as the gravitational wave wavefronts. An important observation is that the single copy has broken gauge invariance due to the mass term proportional to the Ricci scalar. This fact will be discussed at length in our final section. In the following, we will see that other wave solutions give rise to single and zeroth copies that satisfy exactly the same equations.

\subsection{Generalized pp-waves}
Next, we consider the generalization of pp-waves to maximally symmetric curved spacetimes. The case of de Sitter pp-waves is identical to the Kundt waves analyzed above~\cite{Griffiths:2009dfa}; thus, here we will only consider the AdS case. The wavefronts of these AdS waves are hyperboloids that foliate the entire space. The generalized AdS pp-waves are written in Kerr-Schild form with an AdS base metric expressed as
\begin{eqnarray}
\displaystyle \bar g_{\mu\nu} \ud x^\mu \ud x^\nu =\frac{1}{P^2}\left[-2Q^2 \ud u\left(\ud v - \frac{\Lambda}{6}v^2 \ud u\right)+\ud x^2 + \ud y^2\right] \ ,
\end{eqnarray}
with
\begin{equation}
\displaystyle P=1+\frac{\Lambda}{12}(x^2+y^2),\qquad\qquad  Q=1-\frac{\Lambda}{12}(x^2+y^2) \ .
\end{equation}
We choose the corresponding null vector and scalar to be
\begin{align}
k_\mu =e^{-\tanh
	^{-1}\left(P-1\right)}\sqrt{\frac{Q}{P}} \delta_\mu^u \ , \qquad \qquad \quad \phi= e^{2\tanh
	^{-1}\left(P-1\right)} H(u,x,y) \ .
\end{align}
The full Kerr-Schild metric is then a solution to the vacuum Einstein equations provided $H(u,x,y)$ again satisfies \eqref{eq H Kundt}. In the limit $\Lambda\rightarrow0$, this metric reduces to that for pp-waves in flat space~\cite{Griffiths:2009dfa}. We can find the classical copies corresponding to these generalized pp-waves in the same way as in the previous case, and they again turn out to satisfy Eqs. (\ref{AeqW}) and (\ref{SeqW}).

\subsection{Siklos AdS waves}

The Siklos metric in Kerr-Schild form is written with an AdS base metric that reads
\begin{equation}
\bar g_{\mu\nu} \ud x^\mu \ud x^\nu =\frac{\ell^2}{x^2}\left[- 2 \, \ud u  \ud v + \ud x^2 + \ud y^2 \right] \ , \label{adsP}
\end{equation}
where the Kerr-Schild null vector and scalar are chosen to be
\begin{align}
k_\mu =\frac{\ell}{x} \, \delta_\mu^u \ , \qquad \qquad \quad \phi= \frac{x}{\ell} H(u,x,y) \ .
\end{align}
The full metric satisfies the Einstein equations in vacuum provided the function $H(u,x,y)$ is such that
\begin{equation}
\left[ \partial_x^2 + \partial_y^2 - \frac{2}{x^2} \right] H(u,x,y) = 0 \ . \label{eq H Siklos}
\end{equation}

In this case, the wavefronts are planes perpendicular to the $v$ direction. This metric is the only non-trivial vacuum spacetime that is conformal to flat space  pp-waves. In a similar way, one can also construct waves with spherical wavefronts~\cite{Gurses:2012db}. Once again, the single and zeroth copy turn out to satisfy Eqs. (\ref{AeqW}) and (\ref{SeqW}).

\subsection{Shock waves}

\begin{figure}[!b]
	\includegraphics[scale=0.5]{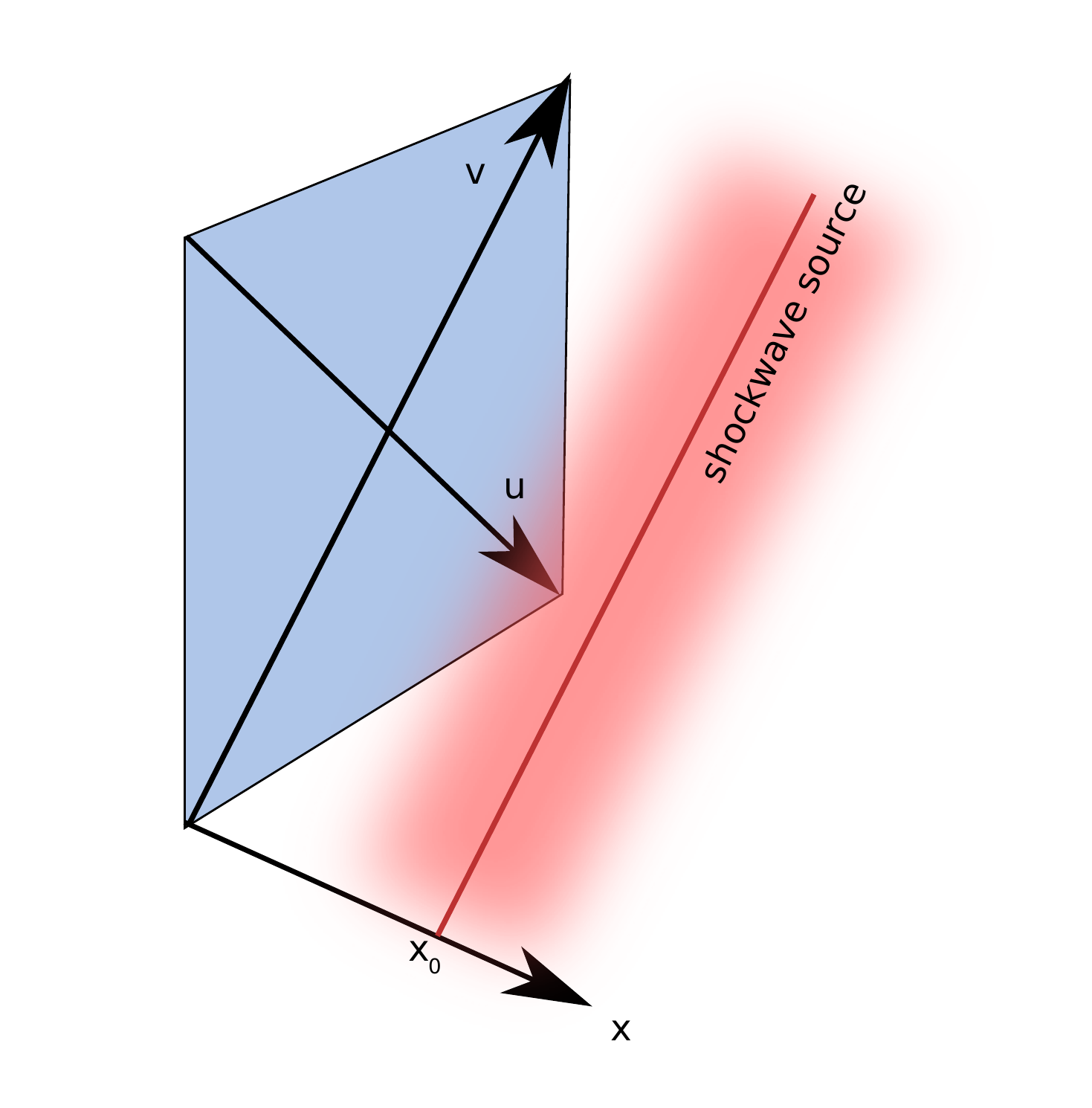}
	\caption{Planar shock wave in AdS. The source travels on a null geodesic at fixed $u=0$, $x=x_0$, and $y=0$. The gravitational, gauge, and scalar shock waves all have this structure.} \label{shocks}
\end{figure}

Finally, we consider planar shock waves in AdS~\cite{Hotta:1992qy,Horowitz:1999gf}. (Note that the case of spherical shock waves follows analogously.) Planar shock waves have the same base metric as Siklos AdS waves---see Eq.~(\ref{adsP})---but unlike the latter they are not vacuum solutions. In this case, the null vector and the scalar field are given by
\begin{align}
k_\mu =\frac{\ell}{x} \delta_\mu^u \ ,\qquad \qquad \quad \phi= \frac{x^2}{\ell^2} H(x,y)\delta(u) \ ,
\end{align}
where we will assume that the source travels on a null geodesic at fixed $u=0$, $x=x_0$, and $y=0$ as shown in figure \ref{shocks}, i.e.
\begin{equation}
T_{\mu\nu} = E\, \frac{x_0^2}{\ell^2}\, \delta(x-x_0) \delta(y)\delta(u)\,\delta_\mu^0 \delta_\nu^0 \ ,
\end{equation}
with $E$  the total energy carried by the shock wave. Notice that we need to place the source away from $x=0$, since the base metric and the Kerr-Schild vector and scalar become singular at that point.  With our ansatz, the Einstein equations reduce to
\begin{equation}
x \left[x \partial^2_x H(x,y)+2 \partial_x H(x,y)+x
\partial^2_y H(x,y)\right]-2 H(x,y)= -16 \pi G E x_0^2 \delta(x-x_0) \delta(y) \ . \label{H eq}
\end{equation}
The solution to this equation is a hypergeometric function, the exact form of which will not be needed here. Imposing  Einstein's equations, the gauge and scalar copies satisfy
\begin{eqnarray}
\nabla_\mu F^{\mu\nu}+\frac{R}{6}A^\nu = g J^\nu \ , \qquad \qquad \quad \nabla^2\phi=j \ ,
\end{eqnarray}
where the sources are
\begin{equation}
J^\nu=j \frac{x}{\ell}\delta^\nu_v,\qquad\qquad \quad  j=2 E \frac{x_0^2}{\ell^2}\frac{x^2}{\ell^2}\delta(u)\delta(x-x_0) \delta(y)\ .
\end{equation}
It is easy to check that the first source follows indeed from Eq. \eqref{current} using the null Killing vector $V^\mu =\delta^\mu_v$. As in the gravitational case, the sources for the shock waves in the gauge and scalar theory are localized at $u=0$, $x=x_0$, and $y=0$.

\section{An unusual example: the BTZ black hole} \label{btz}

Asymptotically flat black holes in $d=3$ space-time dimensions do not exist, but the situation changes in the presence of a negative cosmological constant. Black hole solutions in AdS$_3$ are known as BTZ black holes, and can be viewed as a quotient space of the covering of AdS$_3$ by a discrete group of isometries \cite{Banados:1992wn,Banados:1992gq}. In this section, we construct the single and zeroth copy of these solutions. Given that there are no graviton degrees of freedom in $d=3$, we can at most expect to apply the double copy procedure to the entire BTZ black hole geometry. Therefore, in the following analysis we will use a flat base metric. This approach is different from the one we have adopted in the rest of the paper, since in the previous examples we worked with a curved base metric. This is however an interesting example to consider, because it does not have an immediate counterpart at the level of scattering amplitudes. 

We will write the BTZ black hole metric in Kerr-Schild form with a Minkowski base metric expressed in spheroidal coordinates,
\begin{equation}
\bar g_{\mu\nu} \ud x^\mu \ud x^\nu = -\ud t^2+\frac{r^2}{r^2+a^2}\ud r^2+(r^2+a^2) \ud \theta^2 \ , \label{Mbtz} 
\end{equation}
and a null vector and scalar field given by
\begin{align}
k_\mu =\left(1,\frac{r^2}{r^2+a^2},-a\right) \ , \qquad \qquad \quad \phi =1+8GM+\Lambda r^2\ . 
\end{align}
As for the Kerr black hole, $M$ is the mass of the black hole and $a$ is the angular momentum per unit mass. The corresponding single copy field $A_\mu=k_\mu\phi$ satisfies the Abelian Yang-Mills equations of motion where, as expected, the source is a constant charge density filling all space, that is 
\begin{equation}
J^\mu=4 \rho \delta^\mu_0 \ ,
\end{equation}
where we have replaced the vacuum energy density $\Lambda$ with the charge density $\rho$. By looking at the non-zero components of the field strength tensor $F^{\mu\nu}$,
\begin{equation}
F^{r\,t}=-2\rho\frac{r^2+a^2}{r}  \ , \quad\quad F^{r\,\theta}=-2\rho\frac{a}{r} \ ,
\end{equation}
we can see that the non-rotating case ($a\rightarrow0$) gives rise only to an electric field, whereas the rotating case yields both electric and magnetic fields. Thus, the rotation of the BTZ black hole is translated at the level of the single copy into a non-zero magnetic field, as in the case of the Kerr solution studied in Sec. \ref{kerr}. For completeness, we mention that the equation for the zeroth copy $\phi$ also features a constant source filling all space, i.e. $\nabla^2\phi=-4\rho$.

\section{Discussion and future work} \label{dis}

We have constructed several examples of a classical double copy in curved, maximally symmetric backgrounds. Some black hole copies are straightforward extensions of the double copy in flat space, while other solutions have more involved interpretations. The (A)dS-Schwarzschild and (A)dS-Kerr single copy corresponds to a field sourced by a static and rotating electric charge in (A)dS respectively. Black strings and black branes copy to charged lines and charged planes in (A)dS. A more interesting situation occurs when we consider a black hole in AdS$_3$. The rotating BTZ black hole gives rise to a single copy which produces a magnetic field. Thus, even though there are no gravitons in $d=3$, it seems possible to consider the copy of the geometry. In principle, this relationship should be unrelated to the scattering amplitudes double copy, since there are no gravitons scattering. In this sense, this classical double copy may exhibit a deeper relationship between gravitational and Yang-Mills theories. In all these static cases, the zeroth copy satisfies an equation of motion with a coupling to the Ricci scalar. In $d=4$, it has been conjectured that this is a remnant of the conformal invariance of the Yang-Mills equations. In higher dimensions, it remains unknown if this coefficient is related to a symmetry of these theories. When we turn to time-dependent solutions, the situation seems to change. For the wave solutions, the single copy satisfies an equation of motion corresponding to Maxwell's equation in addition to a term proportional to the Ricci scalar. On the other hand, the zeroth copy equation of motion is simply that of a free scalar field. Despite this change in the equations of motion, we are able to construct the corresponding single and zeroth copies in both time-dependent and time-independent cases.

We have briefly mentioned how some properties (or special limits) of gravitational solutions have no associated mapping to Yang-Mills or scalar fields. This is expected, given that some structures are inherently gravitational, for example horizons. In this sense, when performing the classical copies, one loses information. This is similar to the observation that information of the gauge theory is lost during the BCJ double copy procedure \cite{Bjerrum-Bohr:2013bxa,Oxburgh:2012zr}. Given this, there is no reason to expect that the gravitational instabilities of black hole, black string, or black brane solutions get copied to instabilities in the gauge and scalar theory. Nevertheless, a more detailed study of this should be performed.

Some of our results, obtained by using the classical double copy procedure, are yet to find a completely satisfactory interpretation. One of these is the ambiguity in choosing $k^\mu$ and $\phi$, even after imposing the conditions that $k^\mu$ be geodetic, shear-free, and twist-free. In Sec. \ref{sec:classical double copy} we were able to track down the origin of this ambiguity by extracting Maxwell's equations from the contraction of the Ricci tensor and a Killing vector by using the Einstein equations. In all the examples we have given, we have fixed this ambiguity in a way such that the single and zeroth copies obtained were `reasonable'. Nevertheless, we have yet to identify the exact property required by the null vector and scalar to give rise to the correct copies. This could be related to the fact that, when considering the BCJ double copy, the kinematic factors need to be in BCJ form, where the kinematics factors satisfy the same algebra as the color factors. It is possible that the null vector needs to satisfy a relation that is the analogue of this, but we are not aware of such a relation. We have also found that the time-independent and the time-dependent copies satisfy different equations. For the time-independent case, the scalar copy equation of motion includes an extra factor proportional to the Ricci scalar. In the time-dependent case, this extra factor appears in the equation for the gauge field. These extra factors correspond in both cases to mass terms; this means that the Yang-Mills copy corresponds to a theory with broken gauge symmetry. The reasons for these differences between the stationary and wave solutions remains elusive.

One interesting future direction consists of finding an extension of the Kerr-Schild copy by considering metrics in a non-Kerr-Schild form.  For example, not all waves in $d>4$ can be written in Kerr-Schild form \cite{Ortaggio:2008iq}, but there are examples that can be written in extended Kerr-Schild (xKS) form \cite{Vaidya:1947zz,Ett:2010by}. This xKS form considers the use of a spatial vector orthogonal to the Kerr-Schild null vector. If the Kundt-waves are of Type III, they cannot be written in Kerr-Schild form. Another example of an xKS space time is the charged Chong, Cvetic, Lu, and Pope solution in supergravity \cite{Chong:2005hr,Aliev:2008bh}. Another possible application of this classical copy in curved spacetimes may be in the context of AdS/CFT. The holographic duals to the gravitational AdS solutions that we have considered above have been largely studied in the literature, and it is possible that one could extend the copy procedure to the CFT side of the duality, although this is extremely speculative.\\

\noindent{\bf Acknowledgments}
\\

\noindent We thank Joe Davighi, Matteo Vicino and Adam Solomon for helpful discussions. This work was supported in part by US Department of Energy (HEP) Award DE-SC0013528. R.P. is also supported in part by funds provided by the Center for Particle Cosmology. 

\bibliographystyle{apsrev4-1}
\bibliography{bibliography}

\end{document}